\documentclass[3p,times,twocolumn]{elsarticle}

\usepackage{ecrc}

\volume{00}

\firstpage{1}

\journalname{Nuclear Physics B Proceedings Supplement}

\runauth{}

\jid{nuphbp}

\jnltitlelogo{Nuclear Physics B Proceedings Supplement}

\usepackage{amssymb}
\usepackage{amsmath}

\usepackage[figuresright]{rotating}

\def\snr{SNR~G78.2+2.1}
\def\snrtwo{the $\gamma$ Cygni SNR}
\def\ver{VER~J2019+407}

\def\fgl{1FGL~J2020.0+4049}
\def\fgltwo{2FGL~J2019.1+4040}

\def\psr{PSR~J2021+4026}

\newcommand\veritas{{VERITAS}}

\newcommand\fermi{{\it Fermi LAT}}

\begin{document}

\begin{frontmatter}



\dochead{}

\title{Pulsar Wind Nebulae and Cosmic Rays: A Bedtime Story}


\author{A. Weinstein for the VERITAS Collaboration}

\address{}

\begin{abstract}
The role pulsar wind nebulae play in producing our locally observed cosmic ray spectrum remains murky, yet intriguing.  Pulsar wind nebulae are born and evolve in conjunction with SNRs, which are favored sites of Galactic cosmic ray acceleration.  As a result they frequently complicate interpretation of the gamma-ray emission seen from SNRs.  However, pulsar wind nebulae may also contribute directly to the local cosmic ray spectrum, particularly the leptonic component.  This paper reviews the current thinking on pulsar wind nebulae and their connection to cosmic ray production from an observational perspective.  It also considers how both future technologies and new ways of analyzing existing data can help us to better address the relevant theoretical questions.  A number of key points will be illustrated with recent results from the VHE (E $> 100$ GeV) gamma-ray observatory VERITAS.
\end{abstract}


\end{frontmatter}



\section{Introduction}

The discovery of cosmic rays in the early twentieth century \cite{HESS, Pacini} inaugurated a decades-long, serialized mystery story that has yet to be completed.
At the heart of the mystery lies a set of of simple questions.  What particles appear in the cosmic ray spectrum that we see from Earth?  Where do these particles originate?  How are they accelerated?
In recent decades direct cosmic ray observations have brought us a wealth of information on the cosmic ray spectrum and its composition, clues to the types of environments in which cosmic rays are born.
The interaction of these particles with interstellar magnetic fields, however, prevents them from being traced back to their point of origin.

In order to pursue the other half of this mystery---the nature of cosmic ray accelerators---we must use the secondary photon radiation produced in these cosmic ray nurseries.
These high-energy (300 MeV  - 100 GeV) and very high-energy (VHE; $E>100$ GeV) gamma rays are not deflected by interstellar magnetic fields and can be used to map cosmic ray populations in and near their parent accelerators.
The different processes by which relativistic cosmic ray electrons, protons, and heavier nuclei produce high-energy gamma rays leave
characteristic imprints on the gamma-ray energy spectrum of particular astrophysical accelerators.  These features may be used to constrain the composition and energy spectrum of accelerated particle populations.

The local cosmic ray spectrum is known to be a mix of protons and heavier nuclei, with a smaller admixture of leptons.  A mix of astrophysical accelerators within and outside our Galaxy is believed to contribute, including shocks arising in supernova remnants (SNRs), shocks formed by interacting high-velocity winds from massive stars \cite{2007SSRv..130..439B}, pulsars, and active galactic nuclei (AGN).
We focus here on a single subplot of a much larger story---the role played by pulsar wind nebulae (PWNe) in our quest to understand the Galactic cosmic ray spectrum.

\begin{figure}
  \includegraphics[width=3.2in]{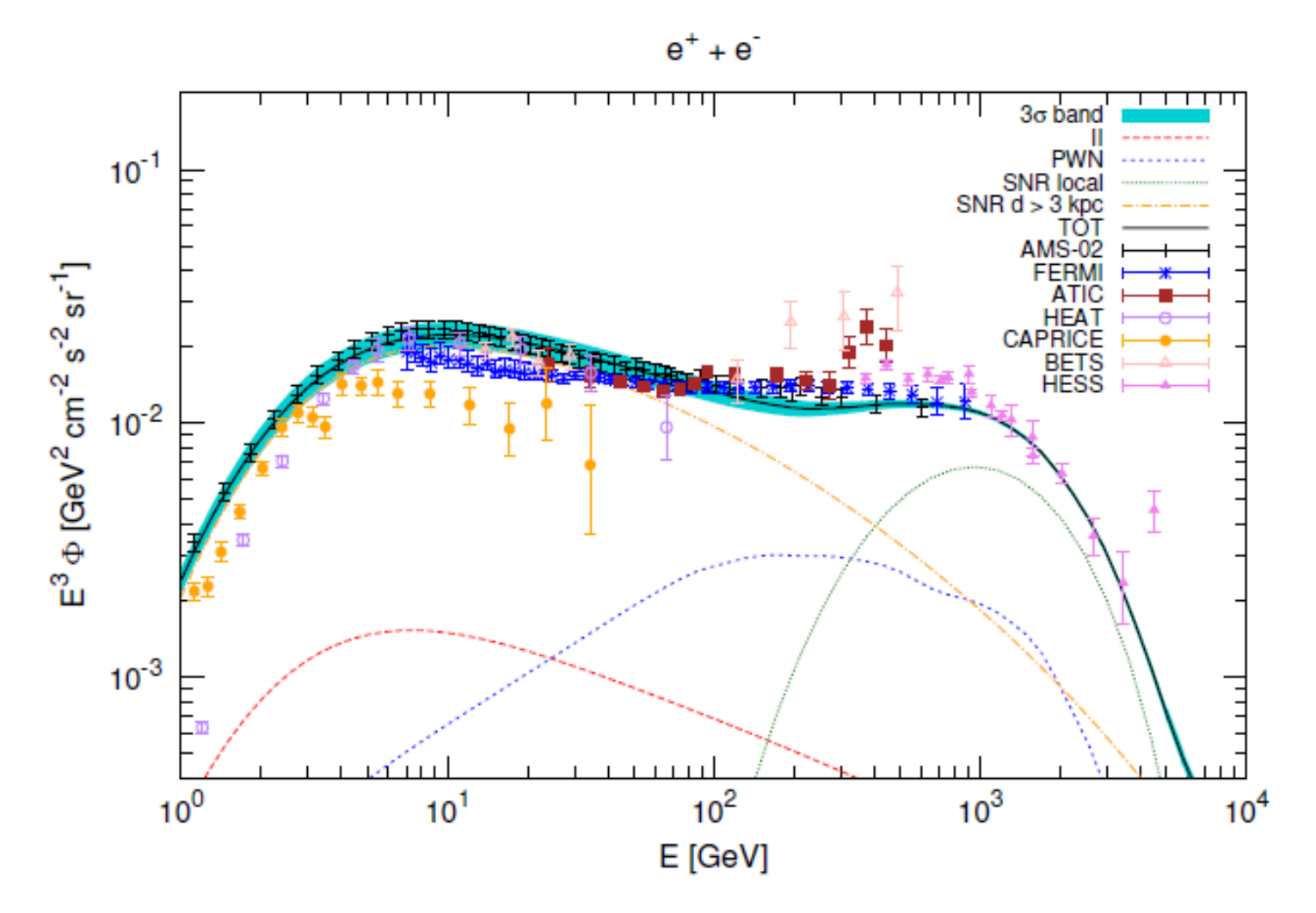}
  \caption{\small Comparison to extant data from AMS-02 \citep{PhysRevLett.110.141102}, \fermi\/ \citep{2012PhRvL.108a1103A}, ATIC \citep{2008Natur.456..362C}, HEAT \citep{1998ApJ...498..779B}, CAPRICE\cite{2001AdSpR..27..669B}, BETS\cite{2008AdSpR..42.1670Y}, and H.E.S.S.\cite{2009A&A...508..561A} of a model considering contributions to the $e^{\pm}$ flux from the following categories: $e^{-}$ from distant ($>3$ kpc) SNRs (dot-dashed yellow) and local SNRs (dotted green), secondary $e^{\pm}$ (long dashed red) and $e^{\pm}$ from PWNe (short dashed blue).  The model is derived from a simultaneous fit to all AMS-02 data. Reproduced from \cite{2014JCAP...04..006D}. (A color version of this figure is available in the online journal).}
  \label{fig:AllLeptonSpectrum}
\end{figure}

\section{Dramatis Personae}

No single gamma-ray observatory provides a complete picture of the gamma-ray sky at all energies and spatial scales.
The Fermi {Gamma}-ray Space Telescope (\fermi\/) can map the entire gamma-ray sky between 300 MeV and 500 GeV over all spatial scales, with an angular resolution that ranges from $0.1^{\circ}$ to $1^{\circ}$ depending on energy.  However, \fermi\/ data become photon-poor above 10 GeV.
Efficient detection of gamma rays above 100 GeV requires ground-based gamma-ray observatories.
Arrays of imaging atmospheric Cherenkov telescopes (IACTs), treat the atmosphere as the sensitive volume of an electromagnetic calorimeter.   Sampling the pool of Cherenkov light produced by atmospheric showers from multiple views permits better reconstruction of the arrival direction and energy of the primary gamma ray \cite{2009ARA&A..47..523H}.  Currently-operating IACT arrays include VERITAS and MAGIC in the northern hemisphere and H.E.S.S. in the southern hemisphere.  These instruments have fields of view between $2^{\circ}$ and $5^{\circ}$.  While the precise energy threshold varies by observatory, all are sensitive between 100 GeV and 10 TeV.
RPC carpet detectors such as ARGO-YBJ \cite{2012ApJ...745L..22B} and water Cherenkov detectors, such as the recently-decommissioned Milagro \cite{2008ApJ...688.1078A,2008PhRvL.101v1101A} and the newly-commissioned HAWC \cite{2013APh....50...26A} have wide fields of view and the best access to energies above 10 TeV, effectively complementing the IACTs.

\subsection{VERITAS}

VERITAS, an array of four 12-m IACTs in southern Arizona, is currently the most sensitive TeV gamma-ray instrument in the northern hemisphere.  Each of the four cameras has a $3.5^{\circ}$-diameter field of view (giving the array an overall field of view of $3.5^{\circ}$) and is equipped with 499-pixel photomultiplier tube camera.  VERITAS has an angular resolution ($68\%$ containment) of better than $0.1^{\circ}$ at 1 TeV and is sensitive to sources with integral fluxes less than $1\%$ of the Crab Nebula flux with less than 30 hours of observation time.
VERITAS has been in continuous operation as a four-telescope array since January 2007.  It underwent a series of upgrades in the period between 2009-2012 that increased the instrument's sensitivity \cite{2011arXiv1111.1225H} and decreased its energy threshold.

\subsection{Pulsar Wind Nebulae}

The rapidly spinning neutron stars known as pulsars are thought to be some of the most powerful sources of electrons and positrons ($e^{\pm}$) in the Galaxy \cite{2014JCAP...04..006D}.  The pulsar is thought to possess a strong, rotating magnetic field that generates a powerful electric field, which then rips particles free from the pulsar surface.  Once accelerated, these charged particles can produce further particle-antiparticle pairs through curvature radiation, creating a relativistic charged-particle wind.  Since the pulsar is created by a supernova explosion, both the pulsar and its wind are initially located within the SNR.  The interaction of the pulsar wind with the slower supernova ejecta creates a \emph{termination shock}.
The magnetized, relativistic plasma between the termination shock and the ejecta is known as the pulsar wind nebula (PWN).  Particles and antiparticles are accelerated to very high energies before becoming trapped by the PWN magnetic field.  Over time the PWN magnetic field weakens and these particles are injected into the interstellar medium (ISM).

\section{Pulsar Wind Nebulae and the Positron Excess}

The origin of cosmic ray electrons and positrons, like the origin of the protons and heavier nuclei that make up the dominant portion of the terrestrial cosmic ray spectrum, is a question of great interest.
Typical models of the cosmic ray lepton spectrum invoke the following contributions, as illustrated in Figure \ref{fig:AllLeptonSpectrum}: electrons produced in SNRs and secondary electrons and positrons from spallation reactions of hadronic cosmic rays with the ISM.  The secondary $e^{\pm}$ components are expected to fall off quickly above 100 GeV; distant SNRs also contribute strongly to the electron spectrum below this energy.  Electrons produced in local SNRs are expected to fill out the electron spectrum above 100 GeV.
In this picture the fraction of positrons in the cosmic ray lepton spectrum should decrease with increasing energy, in contradiction to results from ATIC \cite{2008Natur.456..362C}, PAMELA \cite{2009Natur.458..607A}, \fermi\/ \cite{2009PhRvL.102r1101A} and AMS \cite{PhysRevLett.110.141102}, which show that the positron fraction in fact increases with energy.
While this effect could be produced by WIMP dark matter annihilation, the anomalous positron fraction can also be explained by nearby astrophysical sources, most notably pulsars and PWNe.

$e^{\pm}$ accelerated in the magnetosphere of a mature pulsar are less likely to be confined by an associated PWN. They have consequently been advanced as a viable explanation of the positron fraction excess \cite{2009JCAP...01..025H,2009PhLB..678..283B}.
Most recently, Di Mauro et al. \cite{2014JCAP...04..006D} have used AMS-02 data to constrain pulsar/PWN origin scenarios for the positron excess.  As AMS-02 provides separate measurements of the electron, positron, and all-lepton spectra as well as the positron fraction, these data provide the largest set of simultaneous constraints.  Di Mauro et al. report that these data can accommodate the positron excess using a contribution from either the entire pulsar catalog, a few ($\sim 5$) of the brightest local ($d <1$ kpc) pulsars with ages less than 3000 kyr, or a single local pulsar/PWN.
There are only three single-source candidates for which the fit returns a physically reasonable requirement for the emitted power.  Of these, Di Mauro et al. strongly favor Geminga, a 340 kyr old X-ray and gamma-ray pulsar with a compact ($\sim 2'$) X-ray PWN \cite{2006ApJ...643.1146P}, since it is the only source for which the $e^{\pm}$ emission efficiency is not required to be close to one \cite{2014JCAP...04..006D}.

Others have also noted that the nearby Geminga pulsar could account for the positron excess on its own.  Y{\"u}ksel, Kislev, and Stanev, in particular, point to Milagro's detection of $>20$ TeV gamma-ray emission (MGRO J0634.0+1745) in a $2.6^{\circ}$ by $2.6^{\circ}$ region surrounding the pulsar as evidence for the production, acceleration and escape of $e^{\pm}$ with energies up to 100 TeV \cite{2009PhRvL.103e1101Y}.
These authors note that more detailed constraints on the spectrum and morphology of the TeV source are vital to constraining the particle population.  They also anticipate a gamma-ray halo that extends into the energy range visible to IACTs and note that the halo at these energies provides essential information about the total energetics.  For example, observing the gamma-ray spectrum of the halo soften with increasing distance from the pulsar would provide strong evidence for $e^{\pm}$ cooling.

\begin{figure}[ht]
  \includegraphics[width=3in]{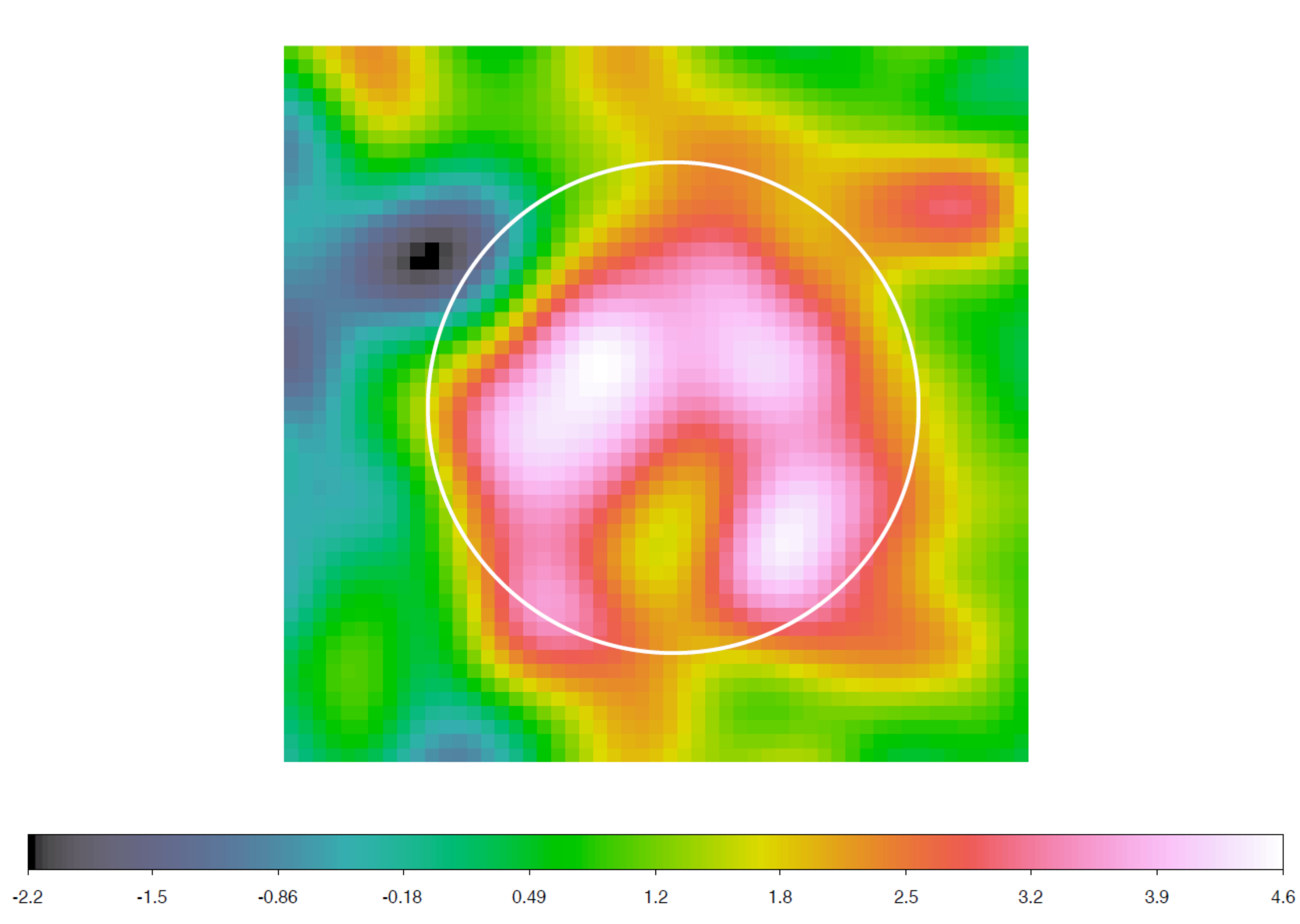}
  \caption{\small Comparison of the VERITAS field of view (white circle), centered at the position of the Geminga pulsar to the morphology of MGRO J0634.0+1745.}
  \label{fig:geminga_fov}
\end{figure}

However, so far our picture of this nebula between X-ray and multi-TeV energies remains blank.  \fermi\/ has only published upper limits on the flux from the PWN \cite{2013ApJ...773...77A}.  VERITAS likewise has no confirmed detection of the Geminga PWN \cite{2008AIPC.1085..269K, 2012PhDT........88F}.
In both cases this gap in our knowledge can be directly attributed to limitations of the instruments and data analysis methods used.
With \fermi\/ data, separating the much weaker PWN emission from that of the extremely bright pulsar is challenging and the methods used to do so---looking at higher energies and looking only at off-pulse emission---reduce the available photon statistics.
In the case of IACT observatories, the main difficulty is illustrated by Figure \ref{fig:geminga_fov}.  Even for VERITAS, which has the most generous field of view of all IACTs in the northern hemisphere, MGRO J0634.0+1745 saturates the field of view.
Standard data-analysis methods such as the \emph{reflected-region model} and \emph{ring-background model} \cite{2007A&A...466.1219B} rely on portions of the field of view distant from the gamma-ray source to estimate the level of irreducible cosmic ray background.  These methods break down when a source fills a large fraction the field of view.

\section{Pulsar Wind Nebulae in a Cosmic Ray Laboratory}

Motivated in part by the hard spectrum of the very high-energy gamma-ray emission seen from the Vela X PWN \cite{2006A&A...448L..43A,2012A&A...548A..38A}, some theorists have advanced hybrid PWNe models, in which the induced electric field extracts both $e^{\pm}$ pairs and heavy nuclei (e.g. iron) from the neutron star surface \cite{2004A&A...423..405B,2007Ap&SS.309..189H,2007Ap&SS.309..179B}.  In this picture, the pulsar wind can play a significant role in accelerating these heavy nuclei and protons resulting from their photo-disintegration, both of which are eventually injected into the interstellar medium \cite{2004A&A...423..405B, 2007Ap&SS.309..179B}.  These models, coupled with the ubiquity of TeV gamma-ray-emitting PWNe \cite{2011ICRC....6..202T,2013ApJ...773..139V}, suggest pulsars/PWNe as a (perhaps the only) source of Galactic cosmic rays above the knee \cite{2004A&A...423..405B, 2007Ap&SS.309..179B}.  This is consistent with measurements that indicate a heavier cosmic ray mass composition above the knee \cite{Haung}.

Moreover, as products of stellar death, pulsars, PWNe, and SNRs are born and evolve in concert.  As a result, PWNe present as a complicating factor when trying to map regions of cosmic ray acceleration and detect signatures of cosmic ray escape.  In the absence of unique spectral signatures, it is often unclear whether high and very high-energy gamma-ray emission should be attributed to particles accelerated within a SNR, or to accelerated leptons in some nearby PWN seen in the radio, X-ray, or optical.  GeV gamma-ray emission from some bright pulsars (particularly those with significant off-pulse emission) can also be difficult to separate completely from the emission produced by a surrounding remnant.

Moreover, no one gamma-ray observatory currently in operation can provide the best determination of a source's gamma-ray spectrum at all energies, nor are these instruments necessarily well-matched in angular resolution.
Milagro in particular has a best angular resolution of $0.5^{\circ}$ at 10 TeV and worse than $1.5^{\circ}$ below 2 TeV.
Consequently, contributions from multiple gamma-ray sources, the majority of which are actual or potential PWNe \cite{2013ApJ...770...93A}, are commingled in key energy ranges with the gamma-ray emission thought to arise from actively or recently accelerated (hadronic) cosmic ray populations.  The inability to separate these sources, particularly at the highest energies, makes it difficult to establish the definitive broadband spectra needed constrain cosmic ray acceleration and escape models.

\begin{figure*}[ht]
  \includegraphics[width=0.85\textwidth]{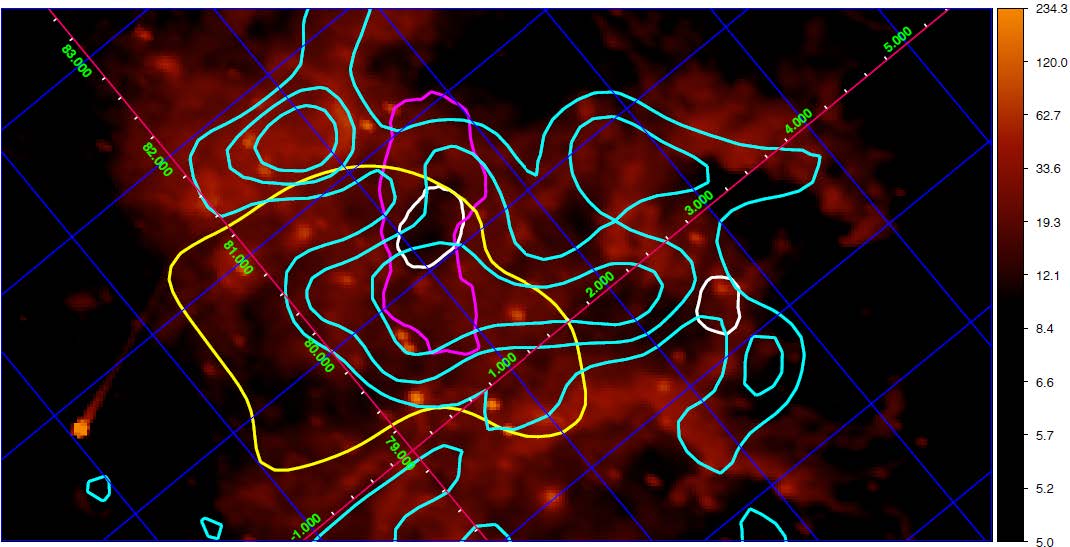}
  \caption{\small MSX 8 $\mu$m infrared survey (color, $\mathrm{W m^{-2} sr^{-1}}$, log scale) for the vicinity of the Cygnus cocoon, showing the cavity in the interstellar medium.
  \fermi\/ 0.16, 0.24 and 0.32 photons $\mathrm{bin^{-1}}$ contours for the cocoon are shown in cyan \cite{Ackermann:2011eq}.
  5 standard deviation contours for MGRO J2031+41, ARGO J2031+41, and the VERITAS sources are shown in yellow, pink, and white respectively.  (A color version of this figure is available in the online journal).
  }\label{fig:cocoon_map}
\end{figure*}

Nowhere are these challenges better illustrated than in the portion of the Galactic Plane that contains the Cygnus-X star-forming region (``Cygnus'').
This region provides a superb laboratory for studying the early phases of the cosmic ray life cycle.  It is the richest star-forming region within 2 kpc of Earth, with a total mass in molecular gas that is at least ten times that of all other nearby star-forming regions combined \cite{2002A&A...392..869L}.
It contains a wealth of massive stars in stellar nurseries, young open clusters, and OB associations, the most massive of which, Cyg OB2, has been the subject of decades of focused observations by gamma-ray observatories \cite{2007ApJ...658.1062K}.  However, the region's very richness lends itself to the types of source confusion discussed above.  Moreover, observations of Cygnus provide a tangential view of the local arm, superimposing structures at many distances along the line of sight \citep{2002A&A...392..869L}.
Conclusively establishing relationships between sources therefore relies on knowing the distance to each source, but these distances are in many cases poorly constrained or entirely unknown.

Two areas of Cygnus are of particular interest.  One region surrounds hard-spectrum, spatially extended gamma-ray emission observed by \fermi\/ above 3 GeV (1FHL J2028.6+4110e) \cite{Ackermann:2011eq, 1FHL} that is interpreted as a cocoon of cosmic rays.  This region also contains Cyg OB2 and the radio and X-ray SNR G78.2+2.1 \cite{2002A&A...392..869L,1977AJ.....82..718H,2000AstL...26...77L}.  The second region surrounds the OB association Cyg OB1.

\subsection{The Cygnus Cocoon and its Vicinity}

The Cygnus cocoon covers an elongated region roughly four square degrees in size and fills a cavity carved in the ISM by stellar winds from a nearby grouping of OB stars.  The hardness of the gamma-ray spectrum between 3 GeV and 500 GeV
indicates that the generating cosmic ray spectrum is also hard, as it would be if the cosmic rays filling the cocoon were freshly accelerated \cite{Ackermann:2011eq,1FHL}.

Determining the precise behavior of the cocoon spectrum at energies above 1 TeV is critical if we are to correctly establish the nature and age of the accelerated particles that fill the cocoon.  The cocoon is co-located with an extended source of $>20$ TeV gamma rays seen by Milagro (MGRO J2031+41) \cite{2009ApJ...700L.127A,2012ApJ...753..159A}.  While it is tempting to use the Milagro result to constrain the behavior of the cocoon spectrum at higher energies, the relationship between these two objects is not so straightforward.
Viewed in gamma rays between a few GeV and 20 TeV, the region surrounding the cocoon is rife with sources, including at least one PWN and a SNR.
As illustrated in Figures \ref{fig:cocoon_map} and \ref{fig:spec_overview}, all or part of the emission from these sources may contribute to the MGRO J2031+41 spectrum.
Moreover, while Ackermann et al. \cite{Ackermann:2011eq} argue that the collective action of winds of massive stars are responsible for accelerating the cocoon particles, contributions from nearby phenomena such as \snr\/ and the Cyg OB2 association cannot be completely ruled out.

\subsection{SNR G78.2+21.1 and VER J2019+407}

\snr\ (the $\gamma$ Cygni SNR) is a nearby ($\sim 1.7 $ kpc) shell-like radio and X-ray SNR $\thicksim\!1^{\circ}$ across ( \citep{1977AJ.....82..718H,2000AstL...26...77L}).  The radio and X-ray emission show higher-intensity features to the north and south \citep{1997A&A...324..641Z,2002ApJ...571..866U}.  At $\sim\!7000$ years \citep{1977AJ.....82..718H,1980A&AS...39..133L,2000AstL...26...77L} it is on the young side of middle age.
It appears to be in an early phase of adiabatic expansion into a low-density medium.

\begin{figure}[ht]
  \includegraphics[width=3.2in]{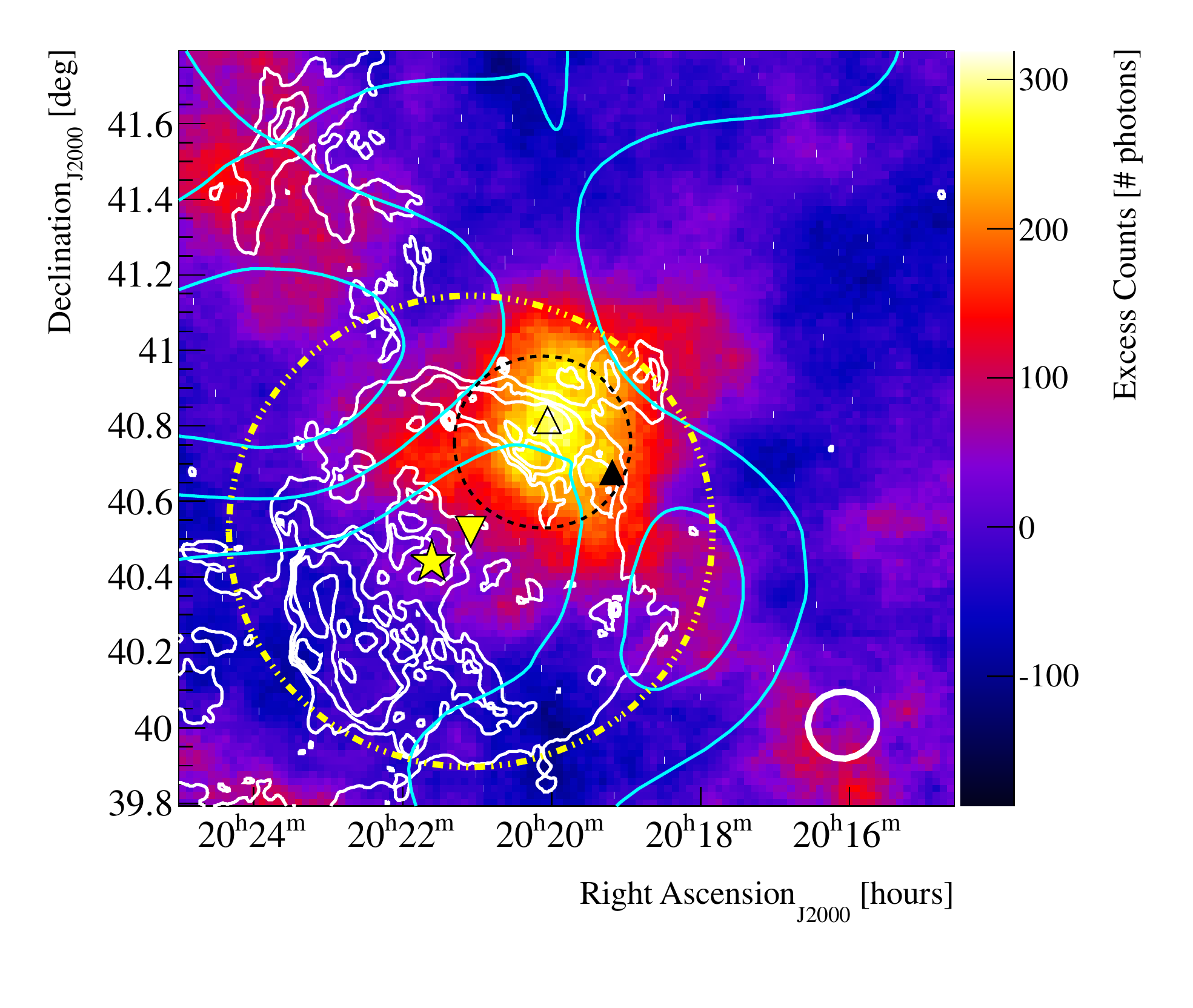}
  \caption{\small Background-subtracted gamma-ray counts map showing the \veritas\ detection of \ver\ and its fitted extent (black dashed circle).  Canadian Galactic Plane Survey (CGPS) 1420 MHz continuum radio contours at brightness temperatures of 23.6K, 33.0K, 39.6K, 50K and 100K (white) \citep{2003AJ....125.3145T} show the extent of the radio remnant.  The star symbol shows the location of \psr. The fitted centroid and extent of the emission detected by \fermi\/ above 10 GeV are indicated by the inverted triangle and dot-dashed circle (yellow).
The open and filled triangles (black) show the positions of the \fermi\/ catalog sources \fgl\ and \fgltwo\/, now subsumed into the extended GeV emission from the entire remnant \citep{2012ApJ...756....5L}.  The 0.16, 0.24, and 0.32 photons $\mathrm{ bin^{-1}}$ contours of the \fermi\/ detection of the Cygnus cocoon are shown in cyan \cite{Ackermann:2011eq}.
The \veritas\ gamma-ray PSF is shown for comparison (white circle, bottom right).  Reproduced from \citep{2013ApJ...770...93A}. (A color version of this figure is available in the online journal).
}\label{fig:verj2019}
\end{figure}

The center of \snrtwo\ hosts a low-luminosity gamma-ray pulsar, \psr\/, which may or may not be the remnant of \snrtwo\/'s progenitor star \citep{2010ApJS..187..460A,2012ApJS..199...31N,2010MNRAS.405.1339T}.  \fermi\/ also observes hard (spectral index $2.39 \pm 0.14$) emission between 10 GeV and 500 GeV from the entire remnant \citep{Ackermann:2011eq,1FHL}.
VERITAS, by contrast, sees a more compact region of emission, \ver\/, coincident with the brightest part of the northern radio shell \citep{2013ApJ...770...93A}.
\ver\/'s nature and relationship to the emission detected by \fermi\/ below 500 GeV from \snrtwo\/ remains unclear. It is highly plausible that \ver\ originates from protons and heavier nuclei accelerated in the SNR shock, but this scenario raises a number of questions.  First, the compactness of \ver\ is puzzling, given that high-intensity radio features are visible in both the north and south and the only indications of molecular material in the portion of the shell opposite to \ver\/.  Furthermore, a reasonable power-law extrapolation of the \fermi\/ source up to 1 TeV suggests that VERITAS should in principle see emission from the majority of the SNR.

This puzzle admits at least two competing physical solutions.
The measured spectra of \ver\ and the emission seen by \fermi\/ are both completely consistent with a power law with $\Gamma=2.4$.  However, the spectra of different regions of \snrtwo\ may evolve differently above 500 GeV, with the emission from the northern shell having a higher-energy cutoff than the remainder of the remnant.  This portion of the remnant would consequently appear brighter at higher energies.
The effect could be accentuated by systematic effects in the data analysis arising from the size of \snrtwo\/ and the presence of a magnitude-2 star overlapping the southeastern portion of the shell.  Such effects would dilute VERITAS' sensitivity to emission from the entire SNR.
\ver\ may also be a chance superposition of a PWN in the \snrtwo\ line of sight.  While this interpretation is not favored, it has recently been lent some weight by the detection of a nearby hard X-ray point source \cite{2013MNRAS.436..968L}.


\subsection{TeV J2032+4130 and Cyg OB2}

The serendipitous detection of the gamma-ray source TeV J2032+4130, based on observations made by the HEGRA IACT system, was reported in 2002 \citep{2002A&A...393L..37A,2005A&A...431..197A}.  Subsequent observations by the IACT observatories MAGIC and VERITAS have confirmed the extension of the source and provided more precise measurements of its spectrum.  All instruments find a power-law spectrum consistent with $\Gamma=2.1$, although there appears to be some disagreement between VERITAS and MAGIC as to the flux normalization \citep{2008ApJ...675L..25A,2014arXiv1401.2828V}.  No measurement thus far has found evidence for either an energy-dependent morphology of the source or a spectral cutoff up to 20 TeV, although current measurements are not highly constraining in either case \citep{2008ApJ...675L..25A,2014arXiv1401.2828V}.

\begin{figure*}[ht]
  \includegraphics[width=0.9\textwidth]{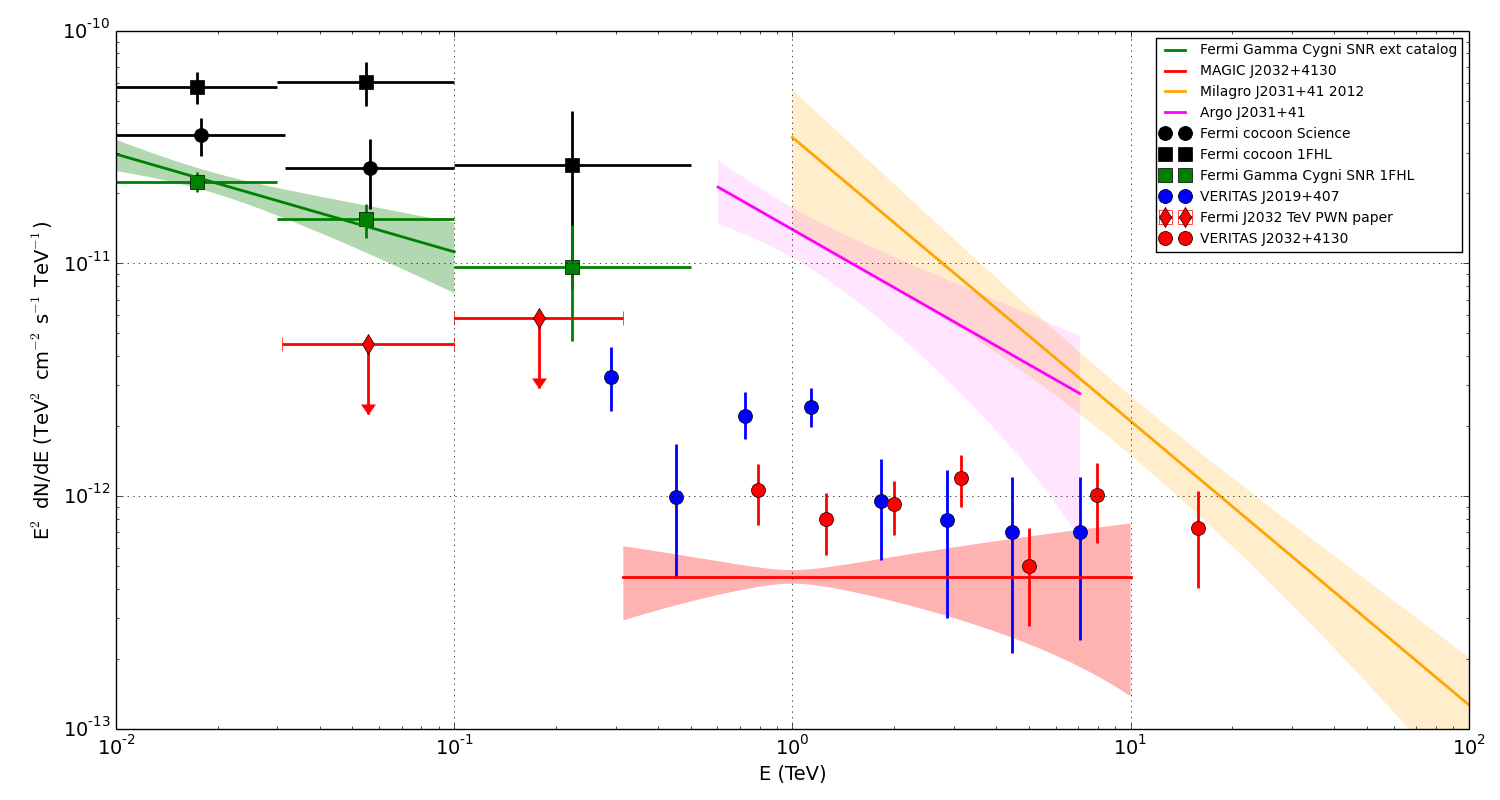}
  \caption{\small Potential contributions to the cocoon broadband spectrum: VER J2019+407 (red dots), TeV J2032+4130 (VERITAS, blue dots; MAGIC, red butterfly), \snrtwo\ (1FHL \citep{1FHL} flux points, green squares; spectrum from \citep{2012ApJ...756....5L}, green butterfly) and MGRO J2032+4130 (yellow butterfly). Also shown: the ARGO J2032+41 spectrum (pink butterfly), the cocoon flux points (\citep{Ackermann:2011eq}, black dots;
  \citep{1FHL}, black squares), and \fermi\/ upper limits on the PWN of PSR J2032+4127 (red arrows). Figure courtesy Luigi Tibaldo.  (A color version of this figure is available in the online journal).}\label{fig:spec_overview}
\end{figure*}

Based on deep VERITAS observations of TeV J2032+4130, Aliu et al. \citep{2014arXiv1401.2828V} strongly suggest that TeV J2032+4130 is a relic PWN powered by the co-located gamma-ray pulsar PSR J2032+4127 \citep{2010ApJS..187..460A}.  Uncertainties in the estimate of the pulsar's distance raise questions about TeV J2032+4130's relationship to other structures in Cygnus-X.
Application of standard Cordes and Lazio \citep{2002astro.ph..7156C} models for dispersion in the Milky Way to recent radio observations \citep{2009ApJ...705....1C} of the pulsar place PSR J2032+4127 at 3.6 kpc, beyond Cyg OB2.  Alternative approaches still place it at 1.7 kpc, consistent with standard distance estimates for both Cyg OB2 and the $\gamma$ Cygni SNR \citep{2009ApJ...705....1C}.
On the grounds that few of the nearby massive OB stars actually overlap the observed VHE gamma-ray emission, Aliu et al. \cite{2014arXiv1401.2828V} disfavor a previously popular interpretation that TeV J2032+4130 is powered by winds from OB stars in the Cyg OB2 association.  However, the stellar wind hypothesis cannot be definitively ruled out.

\subsection{An Incomplete Picture}

Figures \ref{fig:cocoon_map} and \ref{fig:spec_overview} neatly illustrate the challenge of pinning down a broadband spectrum for the Cygnus cocoon from the extant multiwavelength data.  On the one hand, a simple power-law extrapolation of the \fermi\/ spectrum to higher energies would agree with the MGRO J2031+41 spectrum.  On the other, four sources seen below 10 TeV potentially contribute to the emission from MGRO J2031+41: the cocoon, \snrtwo\/, VER J2019+407, and TeV J2032+4130.  There is no question that the cocoon and
TeV J2032+4130 contribute to the emission; the degree to which \snrtwo\ and VER J2019+407 contribute is less clear.
What is clear is that the spectrum of any source contributing to the emission from MGRO J2031+41 must cut off somewhere between 1 TeV and 20 TeV.  Power-law extrapolations of the VER J2019+407 and TeV J2032+4130 spectra would quickly overshoot the MGRO J2031+41 spectrum above 20 TeV; extrapolations of the cocoon and \snrtwo\ spectra would likewise be impossible to accommodate.

Figure \ref{fig:spec_overview} includes the ARGO detection \cite{2012ApJ...745L..22B} for completeness. However, this detection is difficult to reconcile with the other observations as it has a flux level comparable to MGRO J2031+41 but an extent compatible with TeV J2032+4130.

\subsection{MGRO J2019+37 and Cyg OB1}

\begin{figure}[ht]
  \includegraphics[width=3.2in]{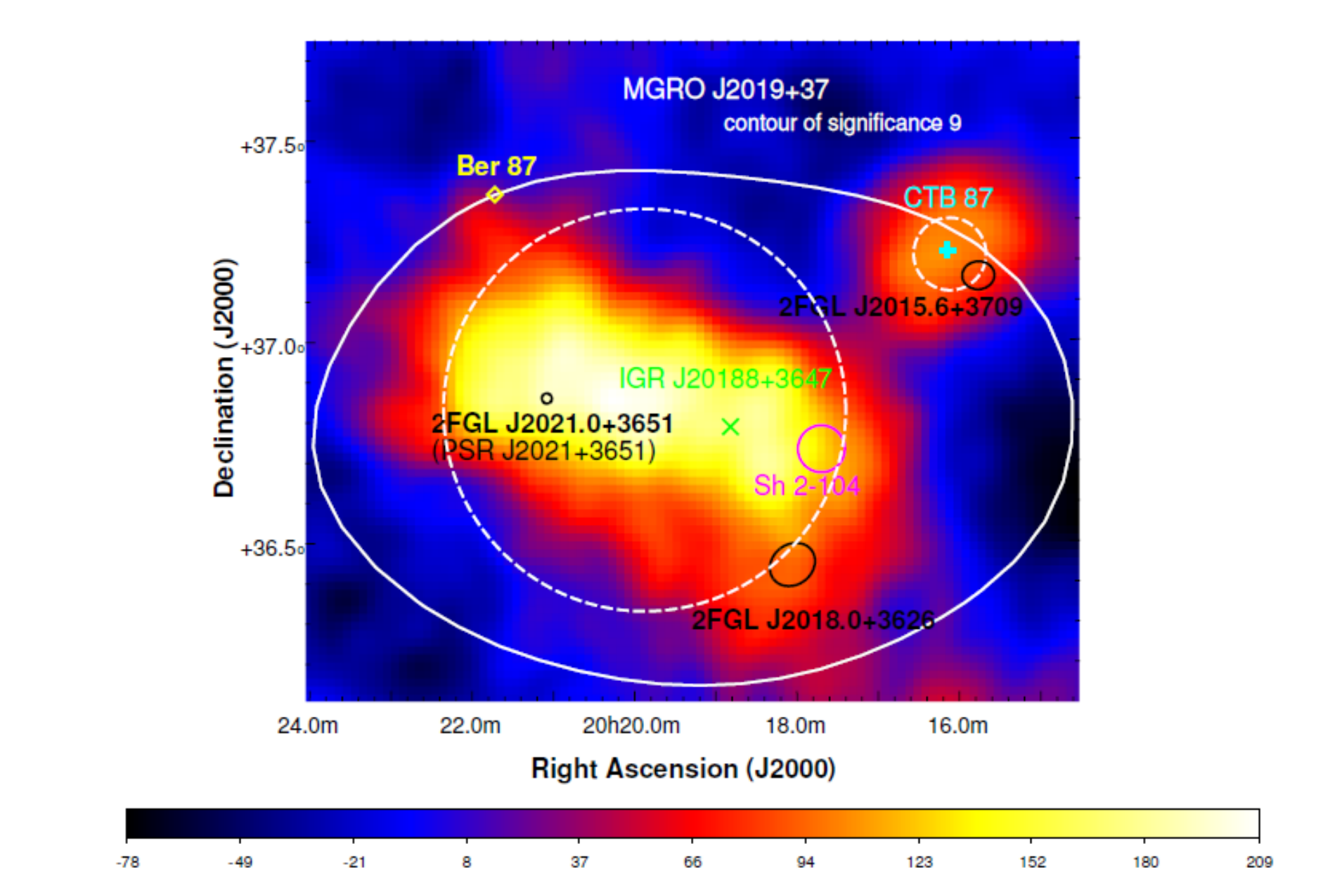}
  \caption{\small Map of the gamma-ray excess above 600 GeV, seen by VERITAS in the vicinity of MGRO J2019+37. The color bar indicates
the number of excess events within a $0.23^{\circ}$ search radius.  White dashed circles indicate the regions used to extract the spectra of VER J2016+371 and VER J2019+368.  The $9\sigma$ significance contour of MGRO J2019+37 is overlaid in solid white.  The remaining solid ellipses, diamonds and crosses indicate the locations of potential counterparts.  Reproduced from \cite{2014arXiv1404.1841V}.  (A color version of this figure is available in the online journal).}\label{fig:MGROJ2019}
\end{figure}

MGRO J2031+41 is not the only extended source seen by Milagro within Cygnus-X.  MGRO J2019+37 is the brightest Milagro source in the Cygnus region, with a flux of about 80\% of the Crab Nebula flux at 20 TeV \cite{2012ApJ...753..159A}.  The bright core of the source has an extent of at least $1^{\circ}$.
As with MGRO J2031+41, indications are that MGRO J2019+37 is a synthesis of multiple gamma-ray sources, including multiple potential PWNe.
Recent observations by VERITAS bear this out.
Figure \ref{fig:MGROJ2019} shows VERITAS' current best picture of MGRO J2019+37 and its vicinity between 600 GeV and 10 TeV \cite{2014arXiv1404.1841V}.  VERITAS already distinguishes two sources in the region: the point-like source VER J2019+368 and VER J2019+368, an apparent ridge of diffuse emission $\sim 1^{\circ}$ in length, roughly bounded by the bright bubble {{\sc H{\thinspace}ii}} region Sh 2-104 to the west and the energetic gamma-ray and radio pulsar PSR J2021+3651 to the east.
There appears to be a clear correspondence between VER J2016+371 and SNR CTB 87.  CTB 87's
radio morphology and the presence of pulsar candidate CXOU J201609.2+371110 within the radio contours support CTB 87's identification as a PWN \cite{2014arXiv1404.1841V}. Aliu et al. \cite{2014arXiv1404.1841V} also argue strongly for a relic PWN interpretation of VER J2016+371/CTB 87.

\begin{figure}[ht]
  \includegraphics[width=3.2in]{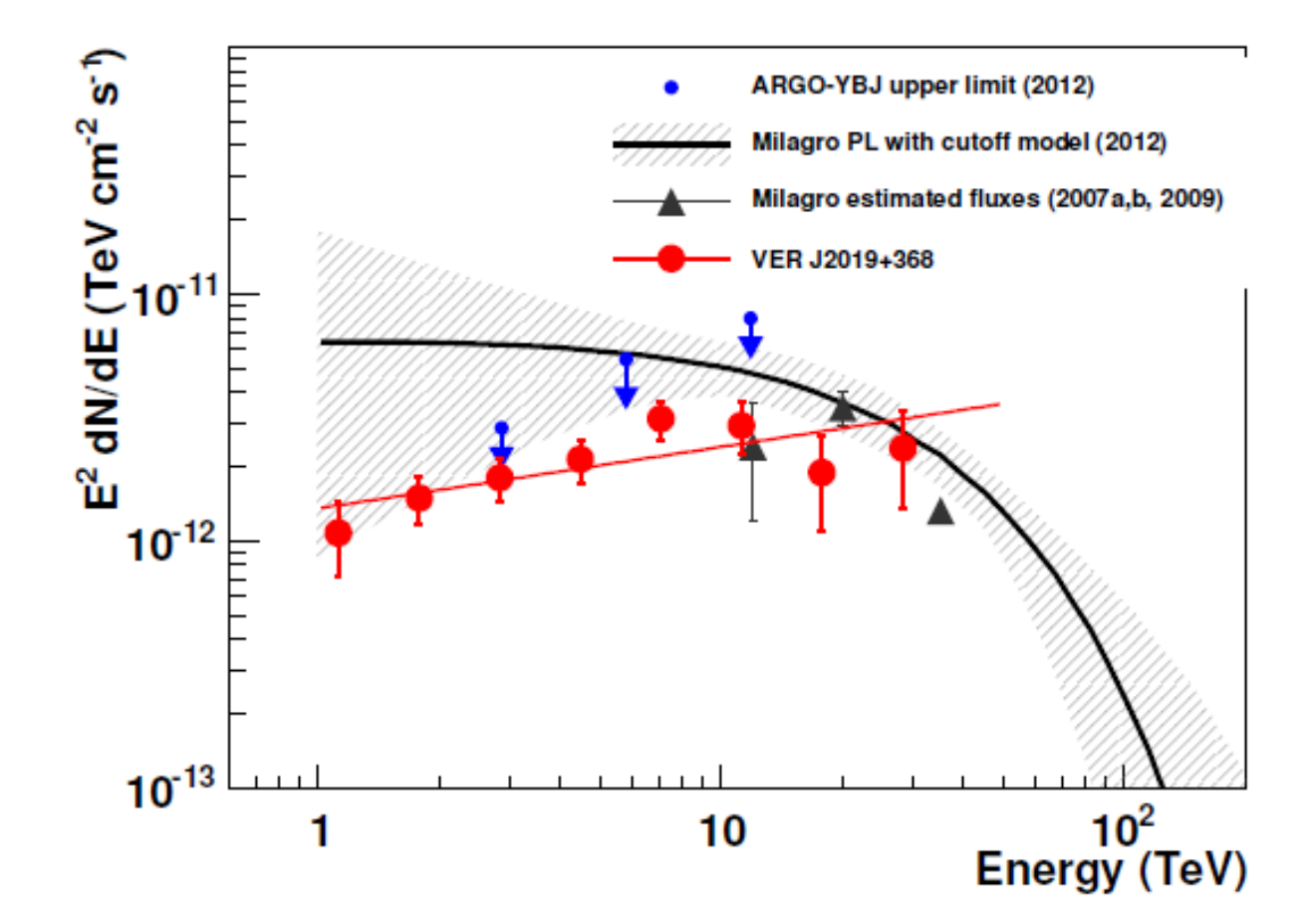}
  \caption{\small Spectral energy distribution of MGRO J2019+37 and VER J2019+368.  The VER J2019+368 spectrum from 1 TeV to almost 30 TeV (red dots) is best fit with a power law with $\Gamma=1.75\pm0.3$.  The original Milagro flux points are shown at 12, 20 and 35 TeV (black triangles) \cite{2007ApJ...658L..33A,2007ApJ...664L..91A,2009ApJ...700L.127A}.  The black curve shows the 2012 best-fit spectrum for MGRO J2019+37: a power law with a cutoff \cite{2012ApJ...753..159A}. The shadowed area corresponds to the 1$\sigma$ band. ARGO-YBJ 90\% confidence-level upper limits for MGRO J2019+37 are shown with blue arrows \cite{2012ApJ...745L..22B}.  Reproduced from \cite{2014arXiv1404.1841V}.  (A color version of this figure is available in the online journal).}\label{fig:MGROJ2019_VERITAS_Spec}
\end{figure}

A unique interpretation of VER J2019+368 and MGRO J2019+37 is more challenging.
Figure \ref{fig:MGROJ2019_VERITAS_Spec} shows that the VER J2019+368 spectrum merges with that of MGRO J2019+37 at high energies.  Taken together with the ARGO-YBJ upper limits, which agree well with the VERITAS spectrum, this argues for VER J2019+368 as the dominant contribution to MGRO J2019+37.
However, VER J2019+368 itself likely incorporates emission from several unresolved sources.  These may include the PWN of PSR J2021+3651 and the {{\sc H{\thinspace}ii}} region Sh 2-104.

\section{The Next Chapter}

In the preceding sections, we considered a number of questions, including Geminga's potential contribution to the positron excess and the nature and origin of the accelerated particles in the Cygnus cocoon, that point towards a common wish list in terms of both new technology and new data analysis techniques.  This wish list includes: improved angular resolution, particularly at multi-TeV energies, to reduce source confusion; improved sensitivity between 100 GeV and 10 TeV to highly extended ($\sim 1^{\circ} - 3^{\circ}$) sources such as \snrtwo\/, the Geminga gamma-ray PWN and the Cygnus cocoon; more effective methods of combining spatial and spectral information from multiple instruments to constrain the accelerated populations responsible for gamma-ray emission; and order of magnitude improvements in detector effective area and sensitivity above 100 GeV.

In this section we primarily consider near-term advances in technology and data analysis methods with the potential to address the first three bullets of this wish list.

\subsection{HAWC}

\begin{figure}
  \includegraphics[width=3.2in]{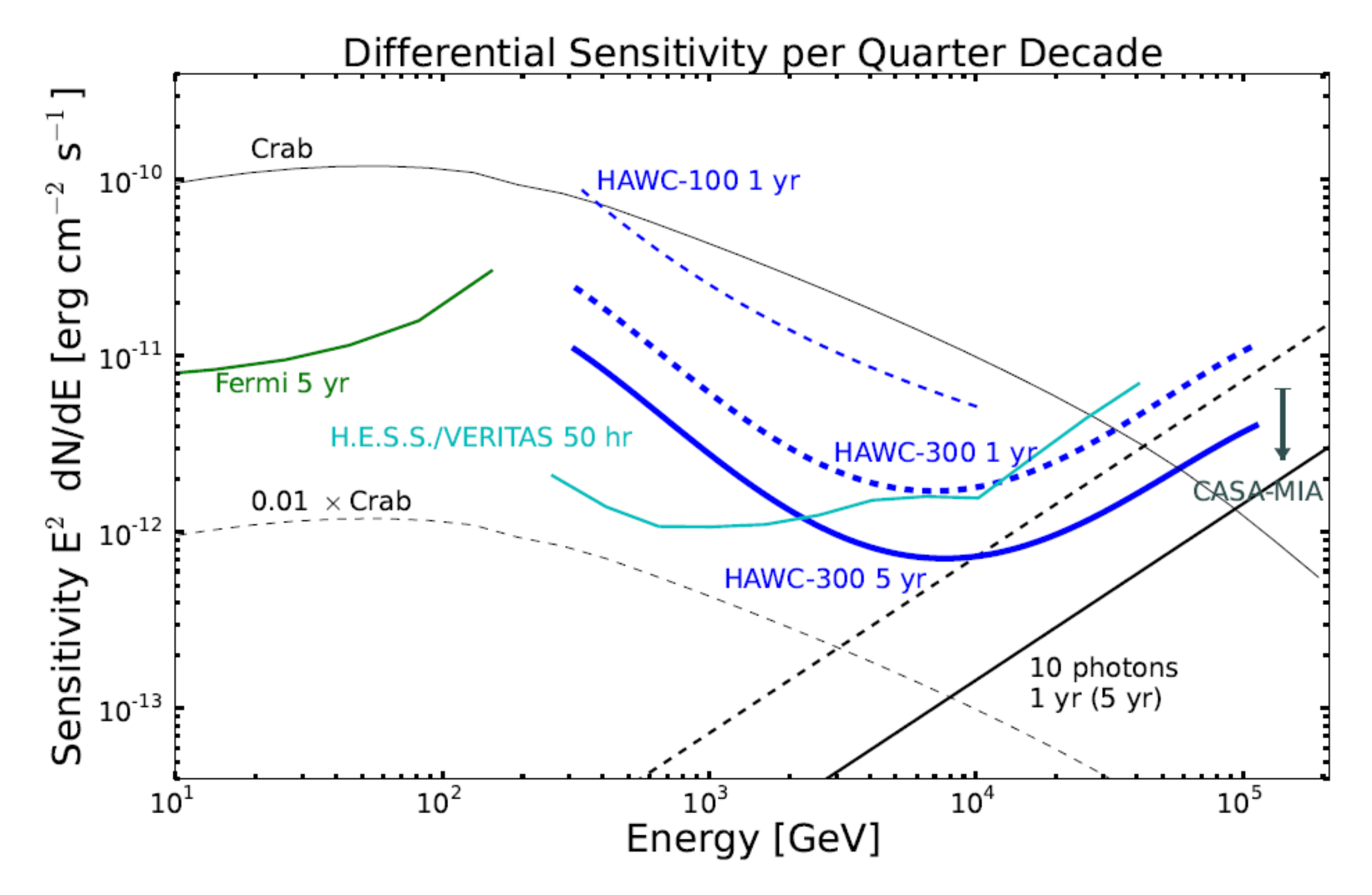}\\
  \caption{\small A comparison of the differential sensitivities of various construction stages of HAWC to those of \fermi\/ and typical IACT instruments.  Reproduced from \cite{2013APh....50...26A}. }\label{fig:diffsens_precta}
\end{figure}

The newly-commissioned High Altitude Water Cherenkov (HAWC) Observatory is a second-generation water Cherenkov instrument, sensitive to both gamma rays and cosmic rays between 100 GeV and 100 TeV \cite{2013APh....50...26A}.  HAWC consists of 300 water Cherenkov detectors, deployed over an area of approximately 22,000 square meters.  Each detector, a water tank with 4 large photomultiplier tubes, samples the energetic secondary particles that reach the ground from an atmospheric shower initiated by a gamma- or cosmic ray.  The HAWC site is 4100m above sea level in the mountains of central Mexico.

Figure \ref{fig:diffsens_precta} shows HAWC's differential sensitivity as a function of energy, compared to both \fermi\/ and current IACT instruments.
Above 10 TeV, HAWC achieves angular resolution ($\sim 0.1^{\circ}$) comparable to that of IACTs at 1 TeV and that of \fermi\/ above 1 GeV.  HAWC, which has angular resolution far superior to that of Milagro at all energies, will bring the sources seen by Milagro into sharper focus over a broader energy range.  However, below 10 TeV HAWC's angular resolution and sensitivity degrade sharply (HAWC's angular resolution is $\sim 0.4^{\circ}$ at 1 TeV).  Since \fermi\/'s effective area begins to degrade past 10 GeV, this creates a ``gap'' in coverage between 100 GeV and 10 TeV.

In the near term, this gap is best filled by IACTs such as VERITAS.  Together, these three instruments can provide unbroken spectral coverage of regions such as Cygnus-X and sources such as Geminga, with similar angular resolution, between 1 GeV and 100 TeV.  In order to take advantage of this treasure trove of data, however, we must develop data analysis methods that allow IACTs to grapple with sources comparable to or larger than the field of view.  We consider such a method in the subsequent section.

\subsection{The (``3D'') Maximum Likelihood Method}

As noted earlier, the most commonly used IACT data analysis methods perform poorly when presented with a source that fills a large fraction of the field of view.
The classic technique of ``on-off'' observation compensates for this by pairing source observations with observations of a blank field, with the off-source field used to estimate the cosmic ray background level.
This method is inefficient as the time spent in off-source observations reduces the available source exposure by at least a factor of two. Unless the ratio of off-source to on-source observations is kept much higher than two, this method also has diluted sensitivity compared to the reflected-region and ring-background methods.  By contrast, an innovative maximum likelihood method can achieve a point-source sensitivity comparable to that of the ring-background model \cite{2013arXiv1308.0055C}.

The unbinned maximum likelihood method discussed here is similar to that used by \fermi\/.  The
``3D'' unbinned maximum likelihood method (3D MLM) proposed for IACT instruments fits a multi-component model to the distribution of photons in both projected sky coordinates and a cosmic ray background rejection parameter.  The inclusion of a background rejection parameter permits the fit to separate gamma-ray source and background components even when the spatial models of these components are indistinguishable \cite{2013arXiv1308.0055C}.
The preliminary studies described in this paper use a well-understood and widely used background rejection parameter, mean scaled width (MSW) \cite{1997APh.....8....1D}.

Probability density functions (PDFs) describe the distribution of the cosmic ray background and any potential gamma-ray sources in all three variables.
The quantity minimized is $-2\ln{\emph{L}}$ where \emph{L} is the product of likelihoods (probabilities of obtaining the observed event data, given the adopted model) over sets of observations distinguished by key characteristics (e.g. the number of telescopes used in event reconstruction, the sky tracking position of the observation, the detector configuration, the zenith angle and the range of reconstructed photon energies).
Each energy bin requires distinct source and background models derived from simulations and cosmic ray data.  Finer energy binning improves sensitivity to spectral parameters but increases the volume of simulations and data needed to develop the models.

\textbf{Models:} The model for a given observation set $i$ (data category) takes the following form for the single-source case:
\begin{equation}\label{eq:principle}
\begin{split}
\mathrm F_i(x,y,w) = {N_i}^{\gamma} (\tau, \alpha_j,  A_{eff}^i, S) S_i^{\gamma} (x,y) \\ \times M_i^{\gamma} (w) + {N_{i}}^{B}(\tau) S_i^{B}(x,y) \times M_i^{B}(w)
\end{split}
\end{equation}
with $S(x,y)$ and $M(w)$ describing the signal and background spatial and MSW distributions, respectively.
The approximation of the three-dimensional PDF by a product of PDFs is valid in the limit where the spatial and MSW distributions are uncorrelated.
The predicted number of events $N_i^{\gamma}$ (the number of source photons in a given data category) is a function of the following quantities: the livetime of the observation, $\tau$, the source spectral parameters $\alpha_j$, and the effective area of the instrument, $A_{eff}$.  $A_{eff}$
describes the probability of detecting a given photon as a function of photon energy and radial offset within the camera, the variations in flux intensity across the surface of the source, the position of the source within the camera, and the fraction of the source contained within the field of view.  It also has a conditional dependence on observation characteristics such as sky brightness, zenith angle, and azimuth angle.

The number and nature of the parameters $\alpha_j$ depend on the source spectral model: e.g. for a power law, $dN/dE=f_0 (E/E_0)^{-\Gamma}$, the $\alpha_j$ would be $f_0$ (differential source flux at a reference energy) and $\Gamma$ (spectral index), while a power-law model with an exponential cutoff would add a third parameter for the cutoff energy.
As the background models are \emph{ad hoc}, derived from cosmic ray data and simulations, the background parameter $N_i^B$ has a relatively straightforward dependence on the livetime, with a conditional dependence on the observation characteristics cited above.  An extended maximum likelihood is used, which accounts for Poisson fluctuations in the total number of events for each category.

\textbf{Simultaneous fitting:} Since the full log-likelihood is merely the sum of the individual log-likelihoods, the data categories may be fit simultaneously.
In a simultaneous fit a separate model, carefully matched in terms of response functions, etc., is used for each data category.  In principle, if each category in the fit had a completely separate set of free parameters, the subdivision of the data into categories would lead to a noticeable loss of statistical power.
However, the source models ultimately depend on the same physical parameters that are used to describe the source morphology and spectrum and these parameters are constrained to be the same across all categories.  For instance, a source modeled as a simple Gaussian in photon arrival direction with a power-law spectrum would have three key parameters: the width of the Gaussian, $f_0$ and $\Gamma$.
This preserves the statistical power of the fit.
The individual background normalization parameters $N^B_i$ may be left as individual free parameters in the fit or constrained to a smaller number of free parameters as appropriate.

\subsection{An Illustrative Test Case}

We illustrate the capabilities of the 3D MLM using simulated observations of the Cygnus cocoon region.
This region contains multiple overlapping sources with radial extensions ranging from a few tenths of a degree to two degrees, all of which are either detected or have the potential to be detected at energies greater than 300 GeV.  Out of all the scientifically interesting regions considered in this paper, the cocoon and environs therefore present the single most complete test of the 3D MLM's capabilities.

The simulations used here are simple toy model simulations based on the expected distribution of both signal and background in the three dimensions (spatial and MSW) of the fit.  They assume a VERITAS-like instrument with comparable sensitivity and identical field of view.  The study uses approximately 200 hours of simulated observations tiled in a non-ideal, irregular exposure pattern.  The version of the 3D MLM used in these studies is a less sophisticated version of the fit described in the preceding section.  In this case the spectral behavior of the sources is not left as a free parameter and less sophisticated models are used for the instrument response functions.  Due to a number of factors, including the time required to develop reasonable background models in each bin of energy, this fit uses a single pair of source and background models, applicable between 500 GeV and 1 TeV.  This focuses the study on an energy range where \fermi\/ provides no data, excludes energies for which the spectral behavior of TeV J2032+4130 is poorly known and excludes energies above 1 TeV where extrapolations of the source spectra grow most uncertain.

The simulation incorporates both of the known gamma-ray sources detected by VERITAS (TeV J2032+4130 and VER J2019+407) and assumes the spectral parameters and spatial extensions reported by VERITAS for these sources.
Simple power-law extrapolations are used for the cocoon and $\gamma$ Cygni, based on the spectral parameters and fluxes reported in \cite{Ackermann:2011eq}, \cite{2012ApJ...756....5L}, and the 1FHL catalog \cite{1FHL}.
The spatial model used in all cases is a two-dimensional Gaussian of the appropriate extent.
The gamma-ray pulsars are not expected to contribute significantly above 500 GeV and are therefore omitted from the simulations.  In these preliminary simulations we have also neglected the contribution from the Galactic diffuse emission.  The Galactic diffuse spectrum falls off rapidly relative to that of any of the sources of interest and the contribution at any given point in the observed region is likely to be below the sensitivity threshold of VERITAS.

\begin{figure}
  \includegraphics[width=3.2in]{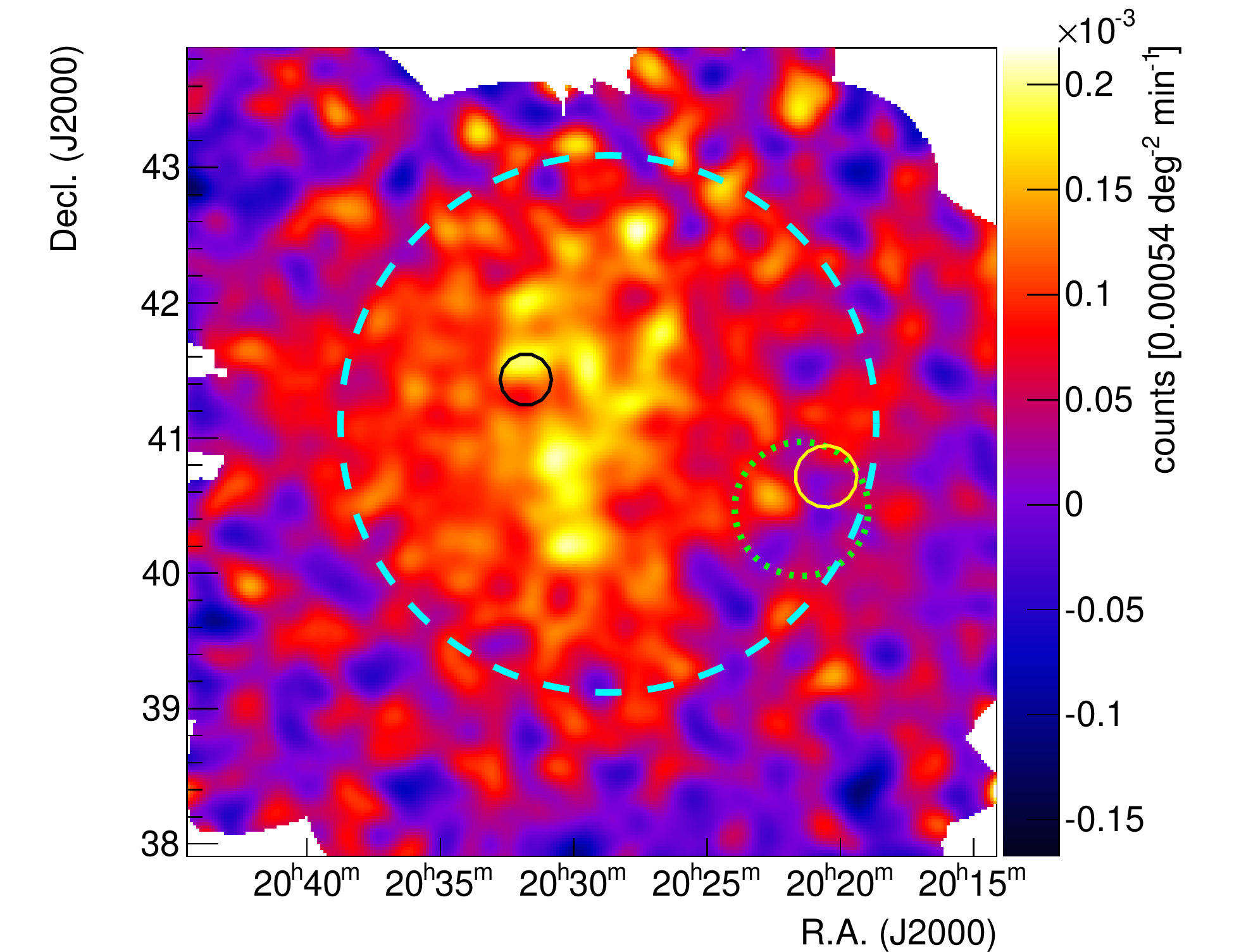}\\
  \caption{\small Smoothed residual map (data minus background model), derived from the 3D MLM applied to a toy model simulation of $\sim 200$ hours of irregularly tiled observations by a VERITAS-like observatory.  1-$\sigma$ contours are shown for the Gaussian source models: the cocoon (dashed blue), $\gamma$ Cygni SNR (short-dashed green), VER J2019+407 (yellow), and TeV J2032+4130 (black).  The latter three sources are included in the background model in this case. (A color version of this figure is available in the online journal).}\label{fig:extrapolations}
\end{figure}

\begin{figure}
  \includegraphics[width=3.2in]{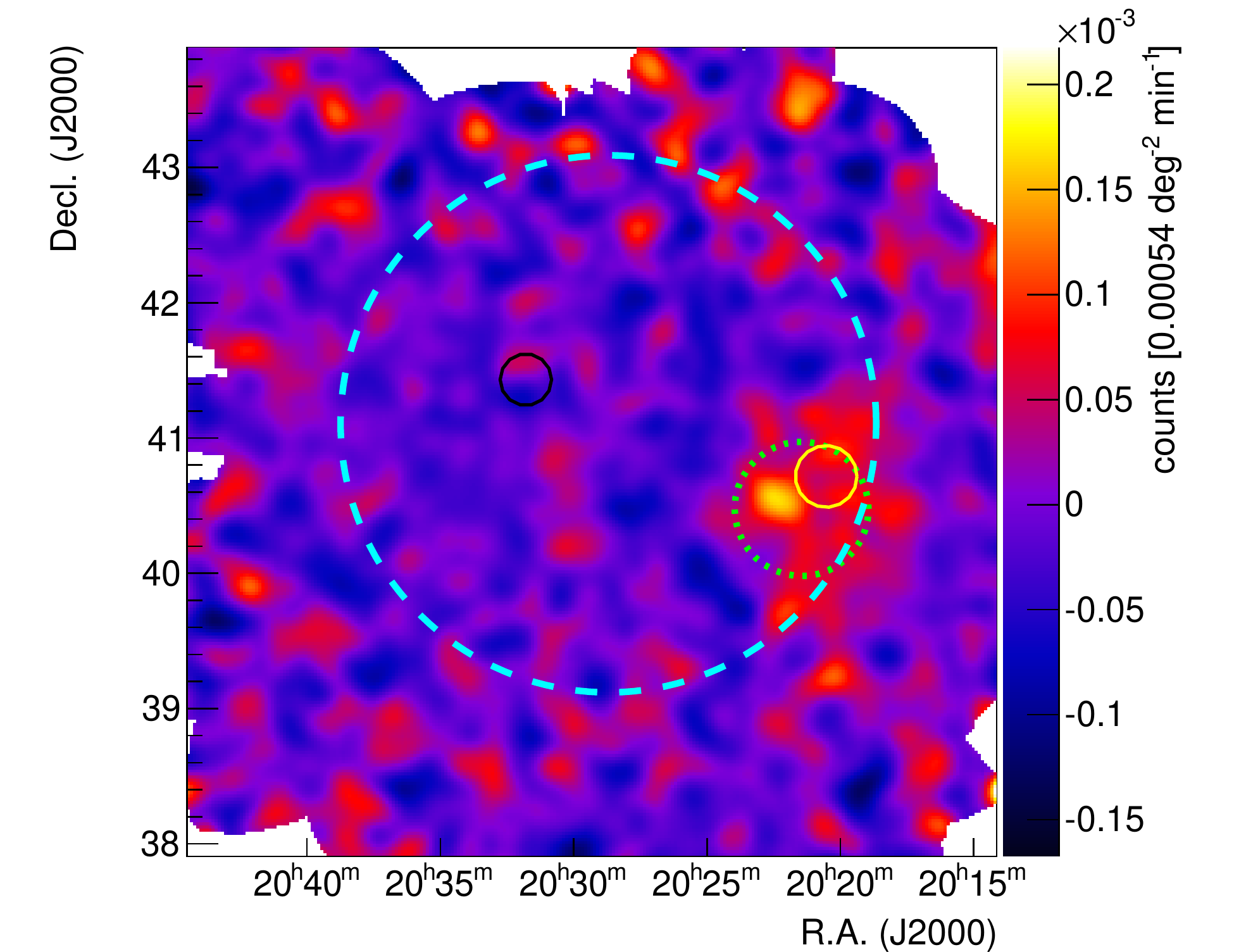}\\
  \caption{\small Smoothed residual map from the same simulated scenario as in Figure \ref{fig:extrapolations}, but with
  all sources but $\gamma$ Cygni included in the background model.  (A color version of this figure is available in the online journal).
   }\label{fig:extrapolations2}
\end{figure}

The square root of the test-statistic, $TS=-2 ln (L_0/L_1)$ where $L_1$ is the maximum likelihood of the test hypothesis and $L_0$ that of the null hypothesis, is used as a proxy for detection significance.
The 3D MLM fit detects emission from the simulated cocoon and simulated $\gamma$ Cygni with $\sqrt{TS}$ values of greater than 45 and 8, respectively.  Figures \ref{fig:extrapolations} and \ref{fig:extrapolations2} also show that the $\gamma$ Cygni and cocoon morphologies emerge clearly.  The fit does a reasonable job of disentangling the overlapping sources of gamma-ray emission, with one exception.
VER J2019+407 can only be distinguished from the $\gamma$ Cygni SNR emission at the 3$\sigma$ level.
It is interesting to note that this particularly challenging case of source confusion parallels the evolution of \fermi\/'s perspective on the SNR.  A source co-located with VER J2019+407, reported in the 1FGL and 2FGL catalogs, was subsequently subsumed into the disk-like emission from the entire remnant \cite{2012ApJ...756....5L}.

It should also be noted that this study makes somewhat pessimistic predictions.  While a simple two-dimensional Gaussian was used to describe the cocoon in \cite{Ackermann:2011eq}, this assumed morphology is almost certainly more featureless and diffuse than the actual cocoon morphology, which should therefore be easier to resolve.  Furthermore, these sources are likely to develop distinct cutoffs at higher energies.  A more complete fit that leaves spectral parameters free and covers the energy range from a few hundred GeV to 10 TeV should provide additional sensitivity and improve the fit's ability to distinguish overlapping sources in this region.

\subsection{Future Instruments}

Future studies of the type discussed here will not be limited to the combined capabilities of \fermi\, IACTs, and HAWC.
The Cherenkov Telescope Array (CTA), a pair of planned next-generation IACT arrays in the northern and southern hemispheres, promises broader energy coverage, dramatically improved ($<0.04^{\circ}$) angular resolution, a comparatively large ($7-8^{\circ}$) field of view and order-of-magnitude increase in sensitivity between 100 GeV and 10 TeV \cite{2009ApJ...698L.133A,2011ExA....32..193A}.
LHAASO, a hybrid instrument combining water Cherenkov and particle detectors \cite{2013SSPMA..43....1C}, also promises unprecedented sensitivity between 20 TeV and a PeV.

\section{Conclusions}

Pulsars and their associated PWNe may play a role in explaining the unexpected increase in cosmic ray positrons at high energies.  In particular, studies of the Geminga PWN at very high energies may constrain the scenario in which a single local source accounts for the entire positron excess.
PWNe also complicate the study of gamma-ray emission from potential cosmic ray nurseries.
Harnessing the combined power of data from \fermi\/, IACTs such as VERITAS, and HAWC to address these questions presents unique but not insurmountable technical challenges.
In the immediate future, combined studies using current-generation instruments and new data analysis techniques will shed new light on the relationship of PWNe to the larger cosmic ray mystery story.
What we learn now will guide even more sensitive studies with powerful next-generation instruments over the coming decades.

\section{Acknowledgments}

A. Weinstein thanks L. Tibaldo, S. Casanova,  L. Grenier, and P. Majumdar for helpful discussions of of the Cygnus cocoon scenario.
 The VERITAS 3D MLM development has been supported in part by NASA grant NNX11A086G.
VERITAS is supported by grants from the U.S. Department of Energy Office of Science, the U.S. National Science Foundation and the Smithsonian Institution, by NSERC in Canada, and by Science Foundation Ireland (SFI 10/RFP/AST2748).  We acknowledge the excellent work of the technical support staff at the Fred Lawrence Whipple Observatory and at the collaborating institutions in the construction and operation of the instrument.




\bibliographystyle{elsarticle-num}
\bibliography{refs3}







\end{document}